\documentclass[preprint,aps,showpacs,showkeys]{revtex4}
\usepackage[dvips]{graphicx}
\usepackage{graphicx}
\usepackage{dcolumn}
\usepackage{bm}

\begin{document}

\title{Lateral shift of the transmitted light beam through a left-handed slab}

\author{Xi Chen}
\email{xchen@graduate.shu.edu.cn} \affiliation{Department of
Physics, Shanghai University, 99 Shangda Road, Shanghai 200436,
People's Republic of China}

\author{Chun-Fang Li}
\email{cfli@staff.shu.edu.cn} \affiliation{Department of Physics,
Shanghai University, 99 Shangda Road, Shanghai 200436, People's
Republic of China} \affiliation{State Key Laboratory of Transient
Optics Technology, Xi'an Institute of Optics and Precision
Mechanics, Academia Sinica, 322 West Youyi Road, Xi'an 710068,
People's Republic of China}


\begin{abstract}

It is reported that when a light beam travels through a slab of
left-handed medium in the air, the lateral shift of the
transmitted beam can be negative as well as positive. The
necessary condition for the lateral shift to be positive is given.
The validity of the stationary-phase approach is demonstrated by
numerical simulations for a Gaussian-shaped beam. A restriction to
the slab's thickness is provided that is necessary for the beam to
retain its profile in the traveling. It is shown that the lateral
shift of the reflected beam is equal to that of the transmitted
beam in the symmetric configuration.


\end{abstract}

\maketitle

\section{Introduction}
The left-handed medium (LHM) with negative permittivity and
negative permeability has attracted much attention
\cite{Smith-1,Smith-2,Shelby,Ziolkowski-1,Parazzoli} and triggered
the debates on the application of the left-handed slab as
so-called ``superlenses"
\cite{Pendry,Hooft,Williams,Valanju,Garcia}. Over 30 years ago,
Veselago \cite{Veselago} first proposed that this peculiar medium
possesses a negative refractive index, which has been demonstrated
at microwave frequencies in recent experiment \cite{Shelby}. In
such media, there are many interesting properties, such as the
reversal of both Doppler effect and Cherenkov radiation
\cite{Veselago}, amplification of evanescent waves \cite{Pendry},
and unusual photon tunneling \cite{Zhang, Liu}. All these
phenomena are rooted in the fact that the phase velocity of light
wave in the LHM is opposite to the velocity of energy flow, that
is, the Poynting vector and the wave vector are antiparallel so
that the wave vector, the electric field and the magnetic field
form a left-handed system. Furthermore, the negative refractive
index has lately been investigated in photonic crystals at optical
frequencies \cite{Notomi,Foteinopoulou}.

It is well known that a totally reflected beam experiences a
longitudinal shift, the so-called Goos-H\"{a}nchen shift, from the
position predicted by the geometrical optics, because each of its
plane wave components undergoes a different phase shift
\cite{Goos}. Recently, P.R. Berman \cite{Berman} and A. Lakhtakia
\cite{Lakhtakia} studied extensively the negative Goos-H\"{a}nchen
shift at an interface between ``normal" and left-handed media. In
order to measure the parameters of left-handed material, I.V.
Shadrivov {\it et al.} \cite{Shadrivov} further investigated giant
Goos-H\"{a}nchen shift in reflection from the left-handed medium.
In addition, J.A. Kong's group \cite{Kong-1} elaborated the
lateral displacement of a Gaussian-shaped beam reflected from a
grounded slab with simultaneously negative permittivity and
permeability. However, the behavior of the transmitted beam didn't
draw as much attention as that of the reflected beam. Only J.A.
Kong {\it et al.} \cite{Kong-2} once discussed the lateral shift
of a Gaussian-shaped beam through a slab of left-handed medium
with the given permittivity and permeability. They concluded that
the displacement is always negative, when the permittivity and
permeability are both negative. More recently, Li and his
co-researchers \cite{Li-1,Huang,Li-2} have investigated the
lateral shift of the transmitted beam through a slab of optically
denser ``normal" medium embedded in the air. It was found that the
lateral shift can be negative, which is similar to the phenomenon
taking place in the LHM \cite{Shelby}. A question arises
naturally: is the lateral shift of the transmitted light beam
through a slab of left-handed medium always negative?

The main purpose of this paper is to report that the lateral shift
of the transmitted beam through a slab of left-handed medium can
be negative as well as positive. The necessary condition is put
forward for the lateral shift to be positive. The positivity of
the lateral shift is closely related to its anomalous dependence
on the slab's thickness, which means that around resonance points,
the absolute value of the negative lateral shift decreases with
increasing the thickness of the slab. It is also shown that the
lateral shift depends on the angle of incidence and the refractive
index. The numerical simulations are performed for a
Gaussian-shaped beam, in order to demonstrate the validity of the
stationary-phase approach. A restriction to the slab's thickness
is obtained that is necessary for the beam to retain its profile
in the traveling. It is pointed out at the same time that the
lateral shift of the reflected beam is equal to that of the
transmitted beam in the simple symmetric configuration. Finally,
we argue the previous opinion that the lateral shift is always
negative when the permittivity and permeability are both negative
and suggest the explanation of the positive lateral shift in term
of the interaction of boundary effects of the slab's two
interfaces with the air.

\section{Lateral shift of the transmitted beam through a left-handed slab}
\begin{figure}[ht]
\includegraphics{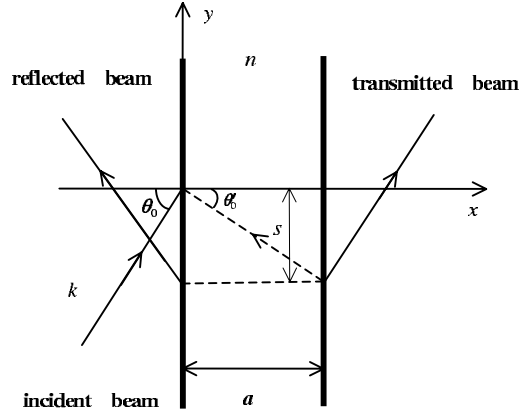}
\caption{\label{fig.1} Schematic diagram of a light beam
propagating through a left-handed slab in the air.}
\end{figure}

For simplicity, we consider a slab of left-handed medium in the
air. Denote by $a$, $\varepsilon$, $\mu$ and $n$, respectively,
the thickness, permittivity, permeability and refractive index of
the slab, extending from $0$ to $a$, as is shown in Fig.
\ref{fig.1}. An incident light beam of angular frequency $\omega$
comes from the left at an incidence angle $\theta_0$ specified by
the inclination of the beam with respect to the x-axis. The field
is assumed to be uniform in the z-direction ($\partial/\partial
z=0$) and time dependence $\exp (-i \omega t)$ is implied and
suppressed. In the case of TE polarization (TM polarization can be
discussed in the same way), the electric field of the plane wave
component of the incident beam is denoted by $
E_{in}(\vec{x})=A\exp(i \vec{k}\cdot \vec{x})$, where $\vec{k}
\equiv (k_x, k_y)=(k\sin\theta, k\cos\theta)$, $k=(\varepsilon_0
\mu_0 \omega^2 )^{1/2}$ is the wave number in the air, and
$\theta$ stands for the incidence angle of the plane wave under
consideration. According to Maxwell's equations and the boundary
conditions, the electric field of the corresponding plane wave of
the transmitted beam is found to be $ E_t(\vec{x}) = F A \exp
\{i[k_x (x-a) +k_y y]\}$, where the amplitude transmission
coefficient $F=e^{i \phi}/f$ is determined by the following
complex number,
$$fe^{i \phi}=\cos k'_x a+ \frac{i}{2}\left(\frac{\chi k_x}{k'_x}+
\frac{k'_x}{\chi k_x}\right)\sin k'_x a,$$ so that
\begin{equation}
\label{phase shift} \phi = \mbox{int}\left(\frac{k'_x
a}{\pi}+\frac{1}{2}
\right)\pi+\tan^{-1}\left[\frac{1}{2}\left(\frac{\chi k_x}{k'_x}+
\frac{k'_x}{\chi k_x} \right)\tan k'_x a\right],
\end{equation}
where $\mbox{int}(\cdot)$ stands for the integer part of involved
number, $k'_x=k'\cos \theta'$, $k'=(\varepsilon \mu
\omega^2)^{1/2}$ is the wave number in the slab, $\theta'$ is
determined by Snell's law, $n\sin\theta'=\sin \theta$, and
$\chi=\mu/\mu_0$. It is clearly seen that the real parameter
$\chi$ can be either positive or negative. When the medium of the
slab is ``normal" material, the parameter $\chi$ is positive. On
the other hand, when the medium of the slab is left-handed
material, the parameter has the property of $\chi<0$.
Correspondingly, the phase shift (\ref{phase shift}) of the
transmitted wave at $x=a$ with respect to the incident wave at
$x=0$ has quite different behavior as compared with that for an
ordinary dielectric slab. Here we are concerned with the lateral
shift of the transmitted beam through a left-handed slab, instead
of a ``normal" dielectric slab.

When measured in the same way as the lateral shift of the
reflected beam as is indicated in Fig.\ref{fig.1}, the lateral
shift of the transmitted beam is defined as $-d\phi/dk_y$
\cite{Artmann,Wigner,Steinberg,Li-3} and is given  by
\begin{equation}
\label{lateral shift} s=\frac{2 \chi k_{y0} a}{k_{x0}}
\frac{k^2_{x0} (k'^2_{x0}+\chi^2 k^2_{x0})- [ k'^4_{x0}+ \chi^2
k^4_{x0}- k^2_{x0} k'^2_{x0}( 1+\chi^2)] \sin2k'_{x0} a/ 2k'_{x0}
a }{4 \chi^2 k^2_{x0} k'^2_{x0} + (k'^2_{x0}-\chi^2k^2_{x0})^2
\sin^2 k'_{x0} a},
\end{equation}
where $k_{x0}=k \cos\theta_0$, $k_{y0}=k \sin\theta_0$,
$k'_{x0}=k'\cos \theta'_0$, and $\theta'_0$ is determined by
Snell's law, $n\sin\theta'_0=\sin \theta_0$. For the case of TM
polarization, the lateral shift $s$ is still valid, if the
parameter $\chi=\mu/\mu_0$ is replaced by
$\chi'=\varepsilon/\varepsilon_0$.

When the permittivity and permeability are chosen to
$\varepsilon=-\varepsilon_0$ and $\mu=-\mu_0$, we have $n=-1$,
$\theta_0=-\theta'_0$, $k_{x_0}=-k'_{x0}$, $k_{y0}=k'_{y0}$, and
$\chi=-1$. In this case, the lateral shift (\ref{lateral shift})
reduces to the following simple form,
\begin{equation}
s=a\tan\theta'_0,
\end{equation}
which is nothing but the lateral shift predicted by the
geometrical optics, the Snell's law of refraction and is in
agreement with the result of J.A. Kong {\it et al.} \cite{Kong-2},
who observed that the lateral shift is always negative when the
permittivity and permeability are both negative. In fact, it will
be shown that the lateral shift $s$ can be positive as well as
negative for a slab of left-handed medium.

\section{Positive property of the lateral shift}
It is seen from the expression (\ref{lateral shift}) for the
lateral shift that when the inequality
\begin{equation}
\label{inequality} k^2_{x0} (k'^2_{x0}+\chi^2 k_{x0}^2)<
[k'^4_{x0} +\chi^2 k^4_{x0} - k^2_{x0} k'^2_{x0}(1+\chi^2 )]
\sin2k'_{x0} a/ 2k'_{x0} a
\end{equation}
holds, the lateral shift is positive for $\chi<0$. It is reversed
in comparison with the prediction of Snell's law of refraction
that the lateral shift of the transmitted beam through a
left-handed slab would be $a\tan \theta'_0$, which is always
negative. Since $ \sin 2k'_{x0} a / 2k'_{x0} a \leq 1$, Eq.
(\ref{inequality}) leads to the necessary condition for the
lateral shift to be positive,
$$k^2_{x0} (k'^2_{x0}+\chi^2 k_{x0}^2)< k'^4_{x0}+ \chi^2 k^4_{x0}
- k'^2_{x0} k^2_{x0}(1+\chi^2),$$ which can be expressed as a
restriction to the incidence angle $\theta_0$ as follow,
\begin{equation}
\label{restriction} \cos\theta_0<\left(
\frac{n^2-1}{1+\chi^2}\right)^{1/2}\equiv \cos\theta_t.
\end{equation}
This means that if the incidence angle satisfies the condition
(\ref{restriction}), that is to say, if $\theta_0$ is larger than
the threshold angle $\theta_t$, one can always find a thickness
$a$ of the slab where the lateral shift of the transmitted beam is
positive. To our surprise, the lateral shift of the transmitted
beam through a left-handed slab is similar to that of the
transmitted beam through an ordinary dielectric slab, predicted by
Snell's law of refraction. In this situation, the positive lateral
shift means the equivalent group index of the slab is positive,
while the phase refractive index is still negative for
$n=-\sqrt{\varepsilon\mu}$. The inequality (\ref{restriction})
shows that positive lateral shifts are more easily implemented at
larger angles of incidence because the larger the angle of
incidence is, the more easily the inequality is satisfied. As a
matter of fact, the inequality (\ref{inequality}) is required for
the lateral shift to be positive. Since the function $\sin2k'_{x0}
a/2k'_{x0} a$ decreases rapidly with increasing $k'_{x0} a$, the
thickness of the slab should be of the order of
$\pi/k'_{x0}=\lambda/[2(n^2-\sin^2\theta_0)^{1/2}]$, so as to make
the positive lateral shift to be significantly large. That is, the
thickness $a$ of the slab should be of the order of the
wavelength, $\lambda$.

\begin{figure}[ht]
\includegraphics{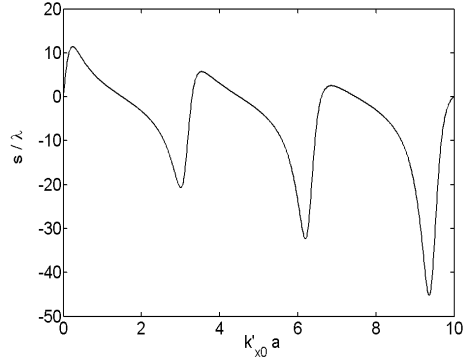}
\caption{\label{fig.2} Dependence of the lateral shift $s$ (in
units of $ \lambda$ ) on the thickness $a$ of the slab, where the
permittivity, permeability and refractive index of the slab are
chosen to be $\varepsilon=-1.89$, $\mu=-0.58$ and $n=-1.05$ at
wavelength $\lambda=23.8$ mm \cite{Parazzoli}, the incidence angle
is $84.5^\circ$, and $a$ is re-scaled by $k'_{x0} a$.}
\end{figure}

A typical dependence of the lateral shift on the slab's thickness
$a$ is shown in Fig. (\ref{fig.2}), where the permittivity,
permeability and refractive index of the slab are
$\varepsilon=-1.89$, $\mu=-0.58$ and $n=-1.05$
($\theta_t=74.0^\circ$) at wavelength $\lambda=23.8$ mm
\cite{Parazzoli}, $a$ is re-scaled by $k'_{x0} a$. In order to
obtain large lateral shifts, a large incidence angle is chosen to
be $\theta_0=84.5^\circ$. Calculation under these conditions shows
that the lateral shift is equal to $67$ mm ($\simeq 2.8 \lambda$)
for $a=40$ mm and is even equal to $232.5$ mm ($\simeq 10
\lambda$) for $a=4$ mm. It is interesting to note that the
oscillation of the lateral shift with respect to $a$ is closely
related to the periodical occurrence of transmission resonance at
$k'_{x0} a=m \pi$ $(m=1,2,3...)$.

Apart from the above-mentioned positivity of the lateral shift
(\ref{lateral shift}) of the transmitted beam, it has other
interesting properties at transmission resonance that deserving
being pointed out. When the transmission resonance occurs, the
transmission probability expressed by
$$
T=\frac{1}{f^2}=\frac{4 \chi^2 k^2_x  k'^2_x }{4 \chi^2 k^2_x
k'^2_x + (k'^2_x-\chi^2 k^2_x)^2 \sin^2 k'_x a}
$$
reaches unity. Thus the light beam is totally transmitted. In this
case, the lateral shift becomes
\begin{equation}
s| _{k'_{x0} a=m\pi}=\frac{k'^2_{x0}+\chi^2 k^2_{x0}}{2 |\chi|
k_{x0} k'_{x0} }a \tan \theta'_0,
\end{equation}
which is negative and less than  $a\tan \theta'_0$ that is
predicted by Snell' law of refraction. Meanwhile, the derivative
of $s$ with respect to the thickness $a$ of the slab is, at
resonance,
\begin{equation}
\frac{\partial s}{\partial a} |_{k'_{x0} a=m \pi}=\frac{k_{y0}}{2
|\chi| k^3_{x0} k'^2_{x0}} \{[k'^4_{x0} +\chi^2 k^4_{x0} -
k^2_{x0} k'^2_{x0}(1+\chi^2 )]-k^2_{x0} (k'^2_{x0}+\chi^2
k_{x0}^2)\} .
\end{equation}
When the condition (\ref{inequality}) is satisfied, this
derivative is more than zero. Therefore, we see that under this
condition the absolute value of the lateral shift decreases with
increasing thickness of the slab around resonance points, because
the lateral shifts around resonance points are negative. In other
words, the positive lateral shift depends anomalously on the
thickness $a$ of the slab around resonance points.

\begin{figure}[ht]
\includegraphics{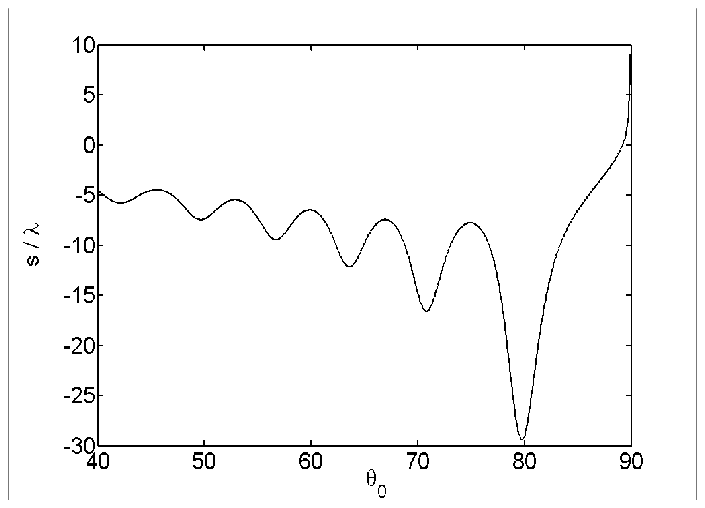}
\caption{\label{fig.3} Dependence of the lateral shift $s$ (in
units of $ \lambda$) on the incident angle $\theta_0$, where the
thickness of the slab $a=6\lambda$, and the other physical
parameters of the left-handed slab are all the same as Fig.
\ref{fig.2}.}
\end{figure}

In addition, it is also indicated from Eq. (\ref{lateral shift})
that the lateral shift depends not only on the thickness $a$ of
the slab, but also on the angle $\theta_0$ of incidence and the
refractive index $n$. To see the latter more clearly, we draw in
Fig. \ref{fig.3} the dependence of the lateral shift on the
incident angle $\theta_0$, where the thickness of the slab is
$a=6\lambda$, and all the other physical parameters are the same
as Fig. \ref{fig.2}. Fig. \ref{fig.3} shows that the lateral shift
decreases with increasing the incidence angle $\theta_0$. It is
seen that the peaks of the lateral shift are approximately
determined by $k'_{x0}a$. That is to say, the oscillation of the
lateral shift with respect to the incidence angle also has close
relation with the periodical occurrence of transmission
resonances. Furthermore, when the incidence angle tends to
$\pi/2$, we have $k_{x0}\rightarrow 0$ and $k'_{x0} \rightarrow
(k'^2-k^2)^{1/2}$. In this limit, the lateral shift takes the
following form,
\begin{equation}
\lim_{\theta_0\rightarrow \pi/2}s= \frac{|\chi| \cot k'_{x0}
a}{k'_{x0} a} a \tan\theta_0,
\end{equation}
which approaches positive infinite. As is shown in Fig.
\ref{fig.3}, a strange phenomenon takes place here that for an
enough large incidence angle ($\theta_0 \rightarrow \pi/2$), the
positive lateral shift becomes larger when increasing the
incidence angle. Of course, the transmission probability $T$ in
this limit tends to zero in the following way,
\begin{equation}
\lim_{\theta_0 \rightarrow \pi/2}T=\frac{4\chi^2 k^2_{x0}}{4\chi^2
k^2_{x0} +(k'^2-k^2)\sin^2k'_{x0} a},
\end{equation}
so that very few light beams can travel through the slab at this
large positive lateral shift.

Now, let us have a brief look at the reflected beam. Denoted by
$RA \exp[i(-k_x x+k_y y)]$ the electric field of the corresponding
plane wave component of the reflected beam, the reflection
coefficient $R$ is determined by Maxwell's equations and the
boundary conditions to be
\begin{equation}
R=\frac{\exp(i\pi/2)}{4f^2}\left(\frac{k'_x}{\chi k_x}-\frac{\chi
k_x}{k'_x}\right)\left[\sin2k'_xa +i\left(\frac{\chi
k_x}{k'_x}+\frac{k'_x}{\chi k_x}\right)\sin^2k'_x a\right].
\end{equation}
The factor that determines the phase of the reflection coefficient
is
\begin{equation}
\label{complex number} \sin2k'_xa +i\left(\frac{\chi
k_x}{k'_x}+\frac{k'_x}{\chi k_x}\right)\sin^2k'_x a.
\end{equation}
If we denote it by $f'\exp(i\phi')$, then the phase of the
reflection coefficient will be $\phi'+\pi/2$. Obviously, we have
$$
\tan\phi'=\tan\phi=\frac{1}{2}\left(\frac{\chi
k_x}{k'_x}+\frac{k'_x}{\chi k_x}\right)\tan k'_x a.
$$
It is meant by this equation that the local properties of $\phi'$
with respect to $k_y$ are the same as those of $\phi$. So the
lateral shift of the reflected beam is locally given by Eq.
(\ref{lateral shift}) \cite{Steinberg,Li-3}. Because $R=0$, at
resonance, $k'_{x0}a=m \pi$, the reflect beam disappears, and its
lateral shift has no definition in this case \cite{Li-2}.  All
these amount to a conclusion that when the resonant transmission
doesn't occur, the lateral shifts of the transmitted and reflected
beam are the same in this symmetric configuration when measured in
the same way.

Of course, when measured with reference to the prediction of
Snell's law, the lateral shift of the transmitted beam will be
$s-a\tan\theta'_0$. Since the lateral shift $a\tan \theta'_0$ is
less than zero, when the lateral shift of the reflected beam is
positive, the lateral shift of the transmitted beam is even
positive with reference to the prediction of Snell's law,
especially at some large angles of incidence.

\begin{figure}[ht]
\includegraphics{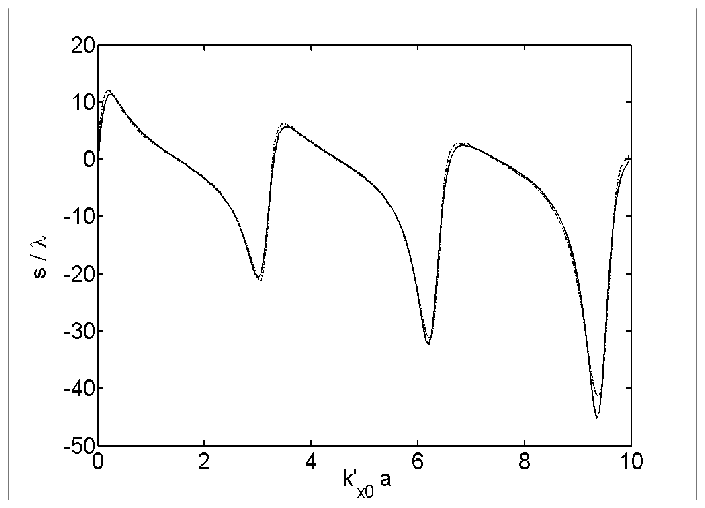}
\caption{\label{fig.4} Comparison of theoretical and numerical
results of lateral shifts (in units of wavelength $\lambda$) with
respect to $a$, where the local waist of Gaussian-shaped beam is
$w=5\lambda$, all the other physical parameters are the same as
Fig. \ref{fig.2}, and $a$ is re-scaled by $k'_{x0} a$. The
theoretical result is shown by the solid curve, and the numerical
results is shown by the dotted curve.}
\end{figure}

To show the validity of the above stationary-phase analysis,
numerical calculations are performed, which confirm our
theoretical results. In the numerical simulation, an incident
Gaussian-shaped light beam is assumed,
$E_{in}(\vec{x})|_{x=0}=\exp(-y^2/2w^2_y+ik_{y0}y)$, which has the
Fourier integral of the following form,
\begin{equation}
\label{incident beam}
E_{in}(\vec{x})|_{x=0}=\frac{1}{\sqrt{2\pi}}\int_{-\infty}^{+\infty}
A(k_y)\exp(i k_y y)dk_y,
\end{equation}
where $w_y=w \sec\theta_0$, $w$ is the local waist of beam, and
the amplitude angular-spectrum distribution is Gaussian,
$A(k_y)=w_y \exp[-(w^2_y/2)(k_y-k_{y0})^2]$. Consequently, the
electric field of the transmitted beam can be written as
\begin{equation}
\label{reflected beam}
E_{t}(\vec{x})=\frac{1}{\sqrt{2\pi}}\int_{-\infty}^{+\infty}
F(k_y)A(k_y)\exp\{i[k_x(x-a)+k_y y]\}dk_y.
\end{equation}
The integral from $-\infty$ to $+\infty$ in Eq. (\ref{incident
beam}) guarantees that the electric field of the incident beam has
a perfect Gaussian profile with respect to $y$. But for a real
incident beam, the incidence angles of its angular-spectrum
components extend from $-\pi/2$ to $\pi/2$. So the integral in the
expression (\ref{reflected beam}) in numerical simulations is
performed from $-k_y$ to $k_y$,
\begin{equation}
E_t^N(\vec{x}) = \frac{1}{\sqrt{2 \pi}} \int_{-k}^{k}
F(k_y)A(k_y)\exp\{i[k_x(x-a)+k_y y]\}dk_y.
\end{equation}
The numerically calculated lateral shift, $s^{N}$, of the
transmitted beam is defined as follows,
\begin{equation}|E_t^N
(x=0,s^N)|^2=\max\{|E_t^N (x=0,y)|^2\}.
\end{equation}
Calculations show that the stationary-phase approximation
(\ref{lateral shift}) for the lateral shift is in good agreement
with the numerical result. In Fig. \ref{fig.4}, we show such
comparisons between the theoretical and numerical results, where
the local waist of the Gaussian-shaped beam is chosen to be
$w=5\lambda$, and all the other optical parameters are the same as
Fig. \ref{fig.2}. It is noted that the discrepancy between
theoretical and numerical results is due to the distortion of the
transmitted light beam, especially when the local waist of the
light beam is narrow. The further numerical simulations show that
the wider the local waist of the incident beam is, the less the
transmitted beam is distorted, and the closer to the theoretical
result the numerical result is.

As pointed out in Ref. \cite{Huang}, for an incident light beam of
the angle spreading $\delta \theta$, the corresponding spreading
of $k'_x a$ should be much smaller than $\pi$, the period of
$|F|$, in order that the stationary-phase approach is valid. This
leads to the following restriction to the thickness of the
left-handed slab,
\begin{equation}
a \ll  \frac{(n^2-\sin^2\theta_0)^{1/2}}{ \delta \theta \sin
2\theta_0} \lambda,
\end{equation}
where $\lambda$ is the wavelength of the light. With the angle
spreading $\delta \theta=\lambda/\pi w$ for a Gaussian-shaped
light beam, we get
\begin{equation}
\label{slab restriction} a \ll  \frac{ \pi
(n^2-\sin^2\theta_0)^{1/2}}{ \sin 2\theta_0} w.
\end{equation}
For instance, if the physical parameters are chosen to be
$n=-1.05$, $\theta_0=84.5^\circ$ and $w=5\lambda$ (corresponding
to the beam divergence of $\delta\theta \sim 4^\circ$), the
requirement (\ref{slab restriction}) is calculated to be $a\ll
11w$. Clearly, this is compatible to the aforementioned
requirement that the slab's thickness should be of the order of
the wavelength, $\lambda$. In addition, it is shown in Fig.
\ref{fig.3} that the angular distance $\Delta \theta_0$ between
two adjacent peaks is determined by $|\Delta k'_{x0}a|=\pi$, which
gives $\Delta \theta_0=\pi/(k_{x0}a \tan\theta'_0)$. In order to
retain the Gaussian-shaped beam's profile in the traveling, the
angular distance $\Delta \theta_0$ should be much smaller than the
divergence of the beam, $\delta \theta$. As a reslut, the
restriction (\ref{slab restriction}) is also required to be
satisfied. In a word, within this restriction the light beam can
travel through the left-handed slab with negligible distortion,
thus the stationary-phase approach in this problem is of validity.

\section{Explanation of the lateral shift and Boundary effects}
The previously discovered lateral shift of the transmitted beam
through a slab of LHM is negative, when the permittivity and
permeability are chosen to be $\varepsilon=-\varepsilon_0$ and
$\mu=-\mu_0$ \cite{Kong-2}. On the basis of this result, the
authors suggested that the lateral shift is always negative when
the medium of the slab is left-handed material. How do we
understand the present positive lateral shift? To this end, we
rewrite the lateral shift (\ref{lateral shift}) as,
\begin{equation}
\label{lateral shift-2} s=\frac{2 \chi k_{x0} k_{y0}
(k'^2_{x0}+\chi^2 k^2_{x0})a}{4 \chi^2 k^2_{x0} k'^2_{x0} +
(k'^2_{x0}-\chi^2k^2_{x0} )^2 \sin^2 k'_{x0} a}+\frac{|\chi|
k_{y0}}{k_{x0} k'_{x0}} \frac{[ k'^4_{x0}+ \chi^2 k^4_{x0}-
k'^2_{x0} k^2_{x0}( 1+\chi^2)]\sin2k'_{x0} a}{4 \chi^2 k^2_{x0}
k'^2_{x0} + (k'^2_{x0}-\chi^2k^2_{x0})^2 \sin^2 k'_{x0} a},
\end{equation}
which consists of two parts. One is a thickness-proportional term
multiplied by a periodical factor with respect to $k'_{x0} a$,
which is always negative for $\chi<0$. The other itself is
periodical. From the lateral shift (\ref{lateral shift-2}), we see
that it is the second term that makes the lateral shift to be
positive. By averaging the two periodical functions over $k'_{x0}
a$ in one period $\pi$, we obtain
\begin{equation}
\overline{s}=a\tan\theta'_0,
\end{equation}
which is always negative. This is exactly what we expect from
Snell's law of refraction. This may be explained as follows.

The negative refraction is inferred from the geometrical optics at
a single interface between the ``normal" and left-handed material,
which has been demonstrated in the experiment \cite{Shelby}.
Moreover, the negative Goos-H\"{a}nchen shift for the totally
reflected beam results from the interaction of the beam with the
single interface of the LHM \cite{Berman,Lakhtakia}. When the
light beam is incident on the left-handed slab at an enough large
angle of incidence, the multiply reflection of the light beam
takes place easily at the slab's two interfaces with the air. This
structure is often analogous to a Fabry-Perot optical
interferometer \cite{Born}: The two interfaces of the slab with
the air play the role of partially transparent mirrors through
which the light is coupled into and out of a resonant cavity. Here
the periodical functions in Eq. (\ref{lateral shift-2}) can be
considered as the result of the interaction of the boundary
effects of the slab's two interfaces, which contribute to the
lateral shift. The averaging over $k'_{x0} a$ just effaces the
interaction, so as to find the geometrical optic prediction.
Actually, the positive lateral shifts presented here can be
understood from the physical viewpoint on the reshaping process of
the light beam by the interference of the multi-reflected beam in
the slab.

When the parameters of the left-handed slab are chosen to be
$\varepsilon=-\varepsilon_0$ and $\mu=-\mu_0$ \cite{Kong-2}, the
negative lateral shift can also be understood by the boundary
effects. In this case, the left-handed medium is a perfect match
to the free space and the interfaces show no reflection, so that
each plane wave component of the light beam can totally travel
through the left-handed slab. Therefore, the lateral shift of the
transmitted beam without reshaping is always negative.
Mathematically, we can know from the Eq. (\ref{lateral shift-2})
that the factors $(k'^2_{x0}-\chi^2k^2_{x0} )^2$ and $[ k'^4_{x0}+
\chi^2 k^4_{x0}- k'^2_{x0} k^2_{x0}(1+\chi^2)]$ are equal to zero
for $\varepsilon=-\varepsilon_0$ and $\mu=-\mu_0$, so that the
periodical functions resulting from the interaction of the
boundary effects don't act on the lateral shift. Therefore, the
opinion that the lateral shifts are always negative when a light
beam travels through a left-handed slab with
$\varepsilon=-\varepsilon_0$ and $\mu=-\mu_0$ is not qualified as
generality. From all these discussions, the lateral shifts of the
transmitted beam through a slab of LHM can be negative as well as
positive, when the permittivity and the permeability are both
negative.

\section{conclusion}
To summarize, we have investigated that the lateral shift of the
transmitted beam though a slab of left-handed medium can be
negative as well as positive by the stationary-phase approach. A
necessary condition (\ref{restriction}) is put forward for the
lateral shift to be positive, which is a restriction to the angle
of incidence. The relation of the lateral shift with its anomalous
dependence on the thickness of the slab around resonances points
is discussed. The lateral shift also depends on the incidence
angle and the negative refractive index. It is shown that the
lateral shift of the reflected beam is equal to that of the
transmitted beam when they are all measured from the normal to the
interface (the $x$ axis) at which the incidence point is located.
In order to demonstrate the validity of the stationary-phase
approach, numerical simulations are made for a Gaussian-shaped
beam. A restriction to the thickness of the slab is obtained that
is necessary for the beam to retain its profile in the traveling.
The positivity of the lateral shift can be explained in terms of
the interaction of the boundary effects of the two slab's
interfaces with the air. Of course, the energy is conserved and
the energy flow can be discussed by the approaches of Lai {\it et
al.} \cite{Lai} and J.A. Kong {\it et al.} \cite{Kong-2}. Finally,
it should be noted that the possibility of observing the positive
lateral shift of the transmitted beam through an ideal, lossless
and nondispersive slab of LHM presented here remains an open
question. However, with the advance of left-handed materials, we
think, the lateral shifts may have potential applications not only
in the measurements of the physical parameters of this material
but also in optical modulations.

\section*{Acknowledgments}

The authors are indebted to Professor V.G. Veselago for providing
his papers. This work was supported in part by the National
Natural Science Foundation of China (Grant No. 60377025), the
Science Foundation of Shanghai Municipal Commission of Education
(Grant No. 01SG46), the Science Foundation of Shanghai Municipal
Commission of Science and Technology (Grant No. 03QMH1405), and by
the Shanghai Leading Academic Discipline Program.

\end{document}